# Understanding the Spin Crossover Dynamical Effects of the Dioxygen Binding and Activation on HOD enzyme

Likai Du[a,b]*

[a]Fundamental Research Center, Xiazkey Information Technology Company, Qingdao, 266035, P. R. China

[b]College of Informatics, Huazhong Agricultural University, Wuhan, 430070, P. R. China

*E-mail: dulikai@xiazkey.tech; dulikai@mail.hzau.edu.cn

**Abstract:**

For the cofactor-free 1-H-3-hydroxy-4-oxoquinaldine-2,4-dioxygenase (HOD), the dioxygen ($O_2$) dependent steps are rate-limiting along with a spin state crossover to the singlet spin state. Here, the primary triplet $O_2$ molecule activation on the 2-methyl-3-hydroxy-4(1H)-quinolone (MHQ) is investigated, and the catalytic role of the intersystem crossing effects is highlighted by directly comparing results from the Born-Oppenheimer dynamics and non-adiabatic surface hopping dynamics. This work confirms non-adiabatic dynamical effects are essential to modulate the $O_2$ activation on the substrate MHQ. The time scale of the equilibration and conversion from triplet to singlet state should be in the range of a few hundreds of femtoseconds. We hope this work provides us a fresh look at the underlying physics of dioxygen activation reactions involving more than one spin state.

**Introduction**

The 1-H-3-hydroxy-4-oxoquinaldine 2,4-dioxygenase (HOD) catalyzes the oxygenolytic breakdown of its natural N-heteroaromatic substrate 2-methyl-3-hydroxy-4(1H)-quinolone (MHQ) with concomitant release of carbon monoxide. Structurally, the HOD enzyme belongs to a particular group of oxygenases, which is known as the α/β-hydrolase fold superfamily of proteins.[1-3] In contrast to other $O_2$ dependent enzymes, the HOD enzyme do not depend on an organic cofactor or a metal ion for catalysis [4]

Recent experimental and theoretical studies have greatly improved our understanding of the catalytic mechanism of HOD enzyme after $O_2$ binding and how it differs from iron-containing dioxygenases.[3,5-8]. And details of the reaction mechanism were established from density functional theory modeling.[8] The reaction is initiated by the triplet O2 molecule and its binding to deprotonated substrate, along with a spin state crossing to the singlet spin state. Note that, the oxygen-dependent steps are rate-limiting as revealed by steady- and transient-state kinetics.[8] Thus, the spin-forbidden $^3O_2$ molecule binds into the substrate MHQ is essential to trigger subsequent redox reaction steps. One of most critical questions in the dioxygen binding process is their non-adiabatic dynamical effects, which requires to reveal why and how the spin state changes actually occur during spin state changes.

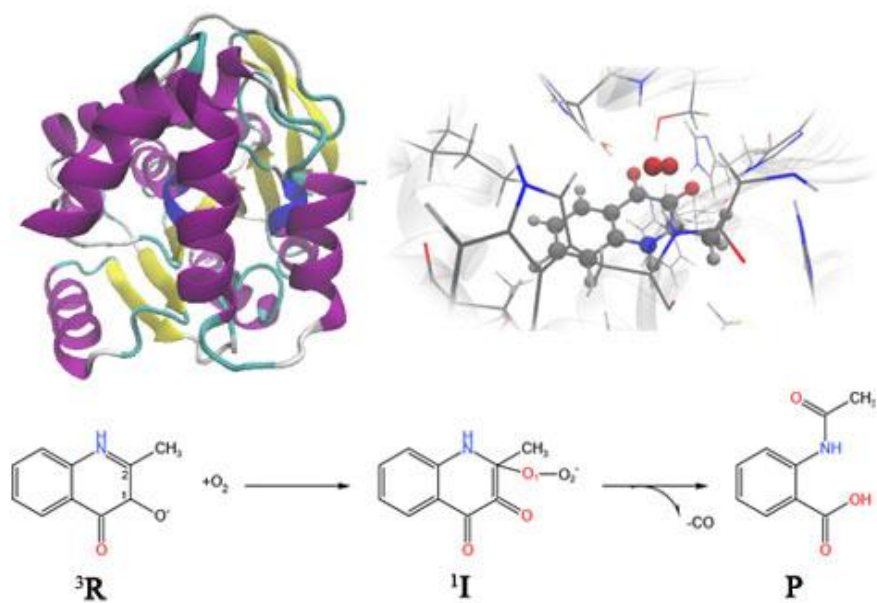

**Scheme 1**. 1-H-3-hydroxy-4-oxoquinaldine 2,4-dioxygenase and the reaction catalyzed by HOD enzyme. The direct reactions between molecular oxygen and organic molecules are spin

forbidden, since molecular oxygen itself has a triplet electronic ground state.

Here, we would discuss the non-adiabatic dynamical effects on the spin forbidden catalytic reaction problems related to the HOD enzyme. Similar as our previous work on the iron and copper model complex, the non-statistical (dynamic) effects of this elementary reaction step are highlighted, for which the intramolecular vibrational energy cannot be fully redistributed in an ultrafast timescale. The non-adiabatic dynamics simulation with coupled singlet and triplet spin electronic states reveals the variation of $O_2$ binding patterns, and we suggest the spin flip events may occur on a few hundreds of femtoseconds under thermal perturbations. Thus, it is interesting to bridge the non-adiabatic dynamics and the catalytic reaction problems, and to reveal how the two-state reactivity mechanism works in a time-resolved manner.

**Theory and Methods**

The $O_2$ binding process in HOD enzyme was studied using the QM and QM/MM methods. For the QM calculations, the molecular geometries of MHQ…$O_2$ complex were optimized at UB3LYP/6-31G(d,p) level, which provides us a simple model system to understand how the spin forbidden reaction works. The QM/MM model was built based on the X-ray crystal structure of HOD enzyme in complexation with the substrate MHQ (PDB ID: 2WJ4 and 2WM2). The QM region includes the organic substrate MHQ, O2, the side chains of His251, Asp126, Ser101, and Trp160 as well as the peptide backbone of Trp36 with the total charge of -1. The hybrid functional B3LYP was selected to describe the QM region, which has been discussed in the previous work. The AMBER ff14SB force field[9] is used for the MM region.

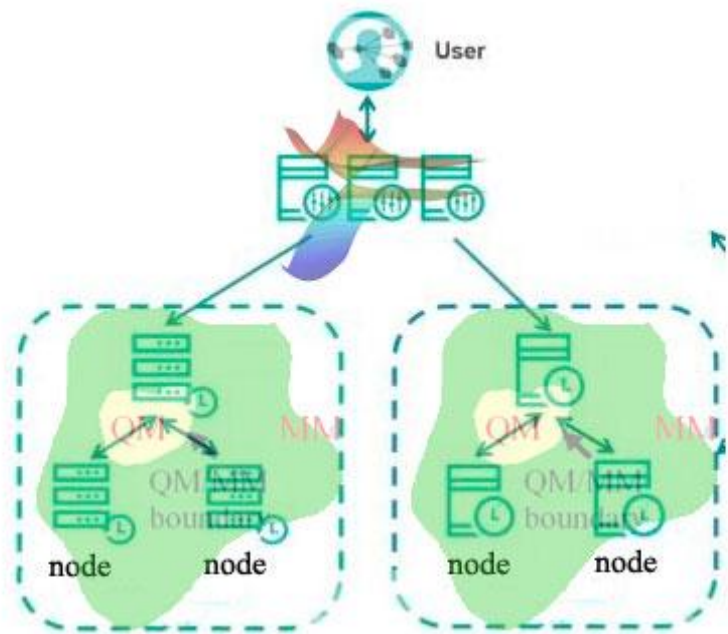

**Scheme 2**. The computational architecture for QM/MM and non-adiabatic dynamics used in this work.

The simulation is most similar to our previous work on the $O_2$ binding iron or copper model complex.[10,11] The on-the-fly non-adiabatic dynamics simulations were done with a modified version of JADE package, which mainly focusing on the non-adiabatic dynamical effects of catalytic problems[10,11], and the machine learning based dynamic models.[12,13] We use the average value between the binding and dissociated geometries, that is 66.1 cm$^{-1}$. This method is shown to give a reasonable result at moderate computational cost. Here, the initial coordinates and momenta were generated for the $O_2$ binding complex, i.e. singlet and triplet minima. The surface hopping dynamics were calculated with 100 trajectories for each initial conditions. The decoherence correction was taken and the parameter is set to α=0.1 Hartree. The dynamic simulations were collected for 500 fs.

**Results and Discussions**

The geometries of the $O_2$ binding HOD enzyme were optimized at their low-lying singlet (S) and triplet (T) states. For the triplet state, the $O_2$ and HQD complex are only loosely connected with the C...O distance of 2.76 Å. For the singlet state, the $O_2$ forms covalent bond with HQD with

C-O distance of 1.55 Å. The singlet state is about 10 kcal higher in energy than the triplet state.

The potential energy surface as a function of the HQD...O2 distance are calculated for the QM/MM and QM model. The potential energy curve undergoes energy crossing between the singlet and triplet state. For the QM model, the crossing geometries show C...O distance of 1.59 Å and O-O distance of 1.34 Å. The MECP is about 10 kcal/mol higher in energy than the stable triplet state geometry. The singlet state PES is generally decay approaching the substrate, while the triplet state PES is unstable when the O2 react with the HQD. This obviously require a spin flip event. The RMSD after overlapping with each other is 0.05 and 0.30 Å for the singlet and triplet states.

To obtain further physical insights about the local minima of the singlet spin state species ($^1$I), we try to monitor the non-adiabatic dynamics starting from the triplet state. The spin population of the triplet state is dominated, and the spin crossover back to the singlet state is minor. Because the triplet state is very high in energy, the sudden excited into the triplet state leads to large kinetic energy and velocities, and the HQD---O$_2$ complex quickly dissociates to be isolated molecules (Figure 1b). In addition, we only observed very few spin flip events in most active trajectories. The dissociation pathway is a roughly representation of the dioxygen binding process.

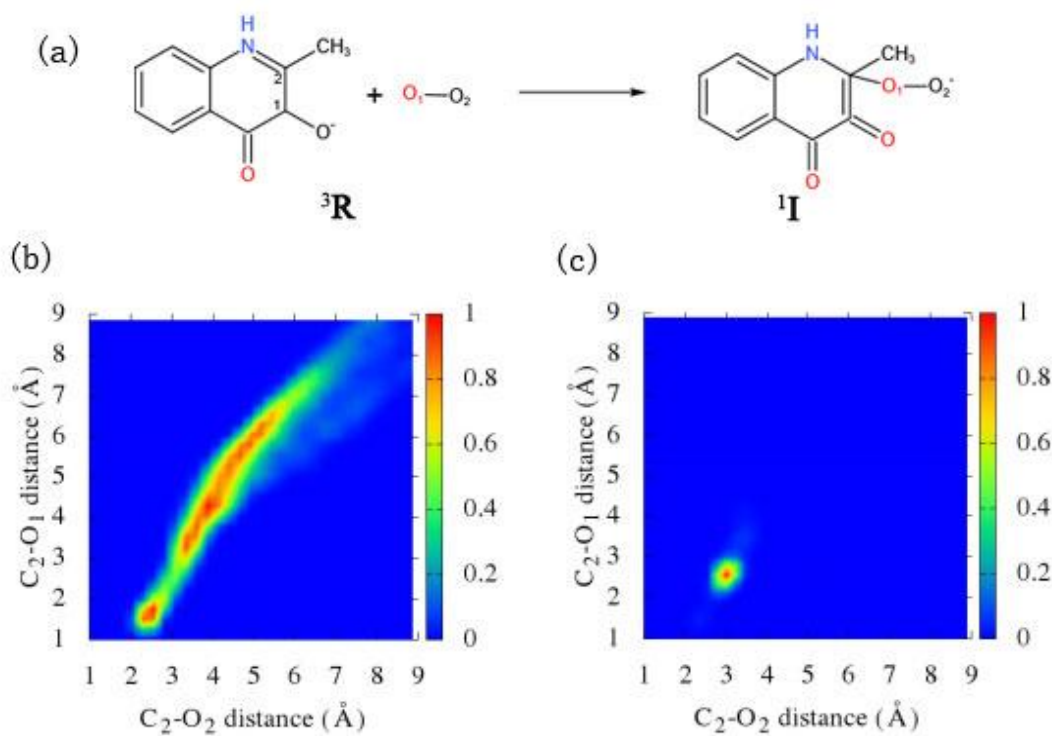

**Figure 1**. (a) The critical species in the dioxygen binding reaction. (b) The distribution of various dioxygen binding modes staring from the triplet state with the local minima structure of the singlet state ($^1$**I**). (c) The distribution of various dioxygen binding modes staring from the singlet state with the local minima structure of the triplet state ($^3$**R**) .

From the local minima of the triplet spin state species ($^3$**R**), we try to perform the non-adiabatic dynamics starting from the open shell singlet state. As shown in Figure 1c, the O$_2$ and HQD complexes are only loosely interacted with the C...O distance of 2~3 Å. The spin flip events are observed in the sub-picoseconds. The population of higher spin state begins to appear after the first 300 fs. And the triplet state reaches to be 20% within 500 fs, and requires more time to reach equilibrium.

The possible dynamics features of the dioxygen binding complexes are shown in Figure 2c. Two primary fates of dynamic trajectories can be observed, one is the dioxygen bounded state ($^1$**I**) and the other is the dioxygen unbounded state ($^3$**R**). The preference for dioxygen bounded states is limited, and hydrogen bond between the oxygen atom and the NH atom is usually preferred (Figure 2d). And thus, the formation of hydroxyl group (-OOH) is possible in a few trajectories. The rebound of the dioxygen molecule is also observed in a few trajectories, which may be enhanced in protein environments, due to their steric effects.

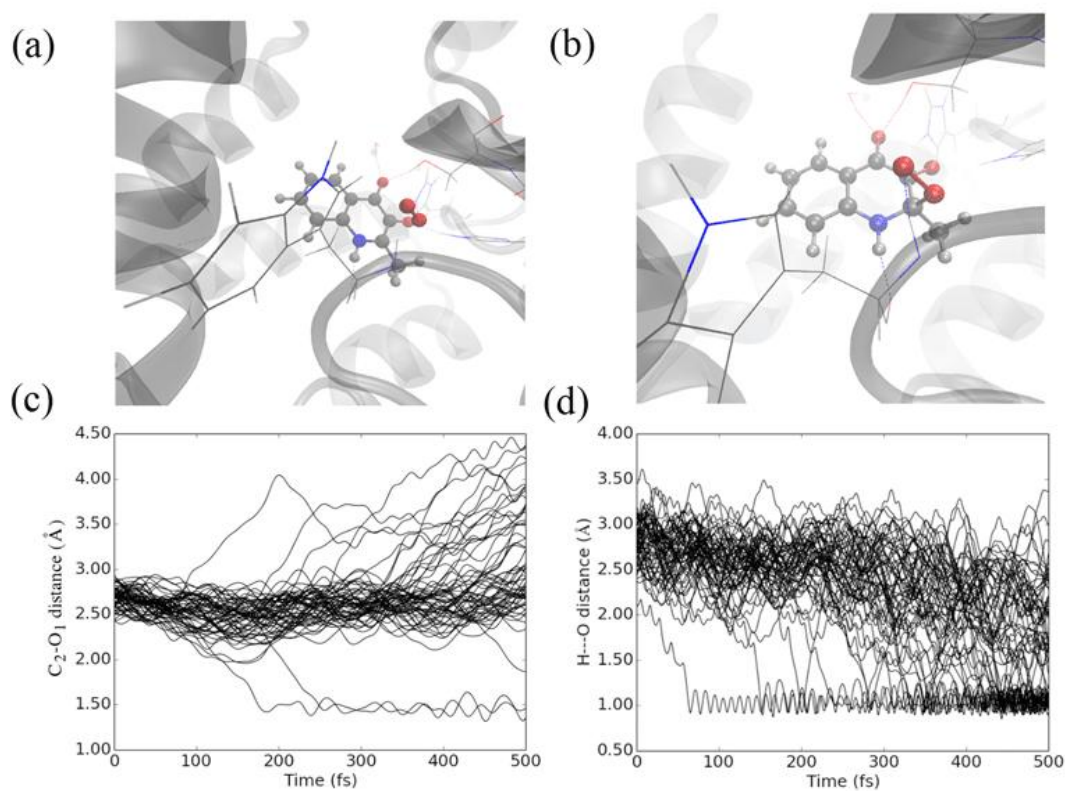

**Figure 2**. The optimized species $^3$**R** (a) and $^1$**I** (b) in the dioxygen binding reaction. The time evolution of C-O bond (c) and O-H$_N$ bond (d) starting from the local minima of the species ($^3$**R**) with singlet spin state.

To explore further physical insights about the dioxygen molecule binding on the substrate, we randomly set up the position and velocities of the O$_2$ molecule near the substrate (Inset in Figure 3a). This initial condition between the dioxygen and substrate may induce a sudden force to drive the relative motion between the dioxygen and substrate. Figure 3a shows the spin states population as a function of time starting from the singlet state. At the beginning, the molecule stays at the singlet state and the population of the triplet state is minor. And the triplet state quickly reaches to be 90% within 200 fs, which is related to the dioxygen unbounded state ($^3$**R**). Figure 3b provides the distribution of C-O distance over dynamic trajectories. And the trajectories are mainly split in two branches in terms of the C-O distance. The oxygen molecule (O$_2$) binding is a critical event to facilitate the subsequent dioxygen activation. As a result, the oxygen atom of species ($^1$**I**) would be much easy to perform its oxidation reactions.

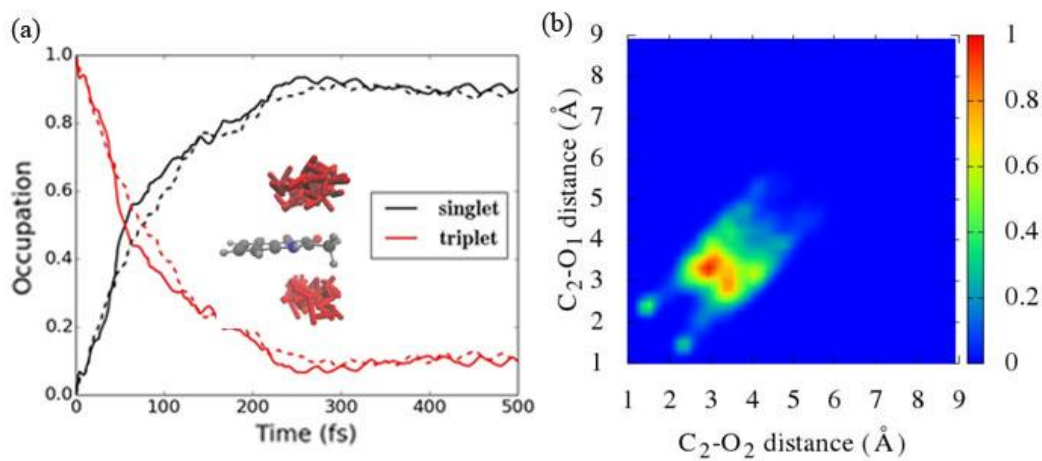

**Figure 3.** (a) The spin state population as a function of time under thermal perturbations. Inset roughly represents the initial geometries for subsequent dynamics. (b) The distribution of various dioxygen binding modes staring from the singlet state with the local minima structure of the triplet state ($^3$**R**) .

**Conclusions**

In summary, the non-adiabatic dynamic simulation provides a new degree of freedom to rationalize spin flip events in the dioxygen activation. Generally, the dioxygen activation process, as a fundamental step in many catalytic reactions, are very important in cellular respiration, corrosion, and industrial chemistry. In this work, we establish the early-stage ultrafast spin flip dynamics of dioxygen binding species. The on-the-fly surface hopping dynamic simulation is applied to obtain this "movie-style" depiction on the coupled low-lying spin state PESs for the HQD enzyme. These results highlight the role of spin flip events on the dioxygen bounded and unbounded states, as a critical factor to determine the catalytic reaction rates. The random thermal fluctuations may easily disrupt established reaction pathways. This provides us insights on numerous possible fates of the dioxygen binding species, as a complimentary for the traditional interpreting of enzymatic reaction mechanism.

**Acknowledges**

The author thanks the support from National Natural Science Foundation of China (No. 21503249). This work was also financially supported by Qingdao Xiazkey Information

**References**

[1] Bauer I,Max N,Fetzner S,Lingens F. 2,4-Dioxygenases Catalyzing N-Heterocyclic-Ring Cleavage and Formation of Carbon Monoxide. Eur. J. Biochem., 1996, 240: 576-583

[2] Fischer F,Künne S,Fetzner S. Bacterial 2,4-Dioxygenases: New Members of the Hydrolase-Fold Superfamily of Enzymes Functionally Related to Serine Hydrolases. J. Bacteriol., 1999, 181: 5725-5733

[3] Steiner R A,Janssen H J,Roversi P,Oakley A J,Fetzner S. Structural basis for cofactor-independent dioxygenation of N-heteroaromatic compounds at the hydrolase fold. Proc. Natl. Acad. Sci. USA, 2010, 107: 657-662

[4] Fetzner S,Steiner R A. Cofactor-independent oxidases and oxygenases. Appl. Microbiol. Biotechnol., 2010, 86: 791-804

[5] Frerichs-Deeken U,Ranguelova K,Kappl R,Hüttermann J,Fetzner S. Dioxygenases without Requirement for Cofactors and Their Chemical Model Reaction: Compulsory Order Ternary Complex Mechanism of 1H-3-Hydroxy-4-oxoquinaldine 2,4-Dioxygenase Involving General Base Catalysis by Histidine 251 and Single-Electron Oxidation of the Substrate Dianion. Biochemistry, 2004, 43: 14485-14499

[6] Hernandez-Ortega A,Quesne M G,Bui S,Heuts D P H M,Steiner R A,Heyes D J,de Visser S P,Scrutton N S. Origin of the Proton-transfer Step in the Cofactor-free 1-H-3-Hydroxy-4-oxoquinaldine 2,4-Dioxygenase: Effect of the Basicity of an Active Site His Residue. J. Biol. Chem., 2014, 289: 8620-8632

[7] Frerichs-Deeken U,Fetzner S. Dioxygenases Without Requirement for Cofactors: Identification of Amino Acid Residues Involved in Substrate Binding and Catalysis, and Testing for Rate-Limiting Steps in the Reaction of 1H-3-Hydroxy-4-oxoquinaldine 2,4-dioxygenase. Curr. Microbiol., 2005, 51: 344-352

[8] Hernández-Ortega A,Quesne M G,Bui S,Heyes D J,Steiner R A,Scrutton N S,de Visser S P. Catalytic Mechanism of Cofactor-Free Dioxygenases and How They Circumvent Spin-Forbidden Oxygenation of Their Substrates. J. Am. Chem. Soc., 2015, 137: 7474-7487

[9] Maier J A,Martinez C,Kasavajhala K,Wickstrom L,Hauser K E,Simmerling C. ff14SB: Improving the Accuracy of Protein Side Chain and Backbone Parameters from ff99SB. J. Chem. Theory Comput., 2015, 11: 3696-3713

[10] Bie L,Liu F,Li Y,Dong T,Gao J,Du L,Yuan Q. Spin crossover dynamics studies on the thermally activated molecular oxygen binding mechanism on a model copper complex. Phys. Chem. Chem. Phys., 2018, 20: 15852-15862

[11] Du L,Liu F,Li Y,Yang Z,Zhang Q,Zhu C,Gao J. Dioxygen Activation by Iron Complexes: The Catalytic Role of Intersystem Crossing Dynamics for a Heme-Related Model. The Journal of Physical Chemistry C, 2018, 122: 2821-2831

[12] Du L,Lan Z. An On-the-Fly Surface-Hopping Program JADE for Nonadiabatic Molecular Dynamics of Polyatomic Systems: Implementation and Applications. J. Chem. Theory Comput., 2015, 11: 1360

[13] Liu F,Du L,Zhang D,Gao J. Direct Learning Hidden Excited State Interaction Patterns from ab initio Dynamics and Its Implication as Alternative Molecular Mechanism Models. Sci. Rep., 2017, 7: 8737